\documentclass[twopage,11pt]{article}
\usepackage{times}
\usepackage{graphicx}
\usepackage{fancyhdr}
\usepackage{amssymb}
\usepackage{amsmath}
\usepackage{fancybox}
\usepackage{epsfig,exscale,psfrag,color,cite}

\begin{document}

\title{Ferromagnetic phase transition in spinor Bose gases}

\author{Qiang Gu\\
Institut f\"ur Laser-Physik, Universit\"at Hamburg,\\
Luruper Chaussee 149, 22761 Hamburg, Germany}

\date{}
\maketitle                            

\begin{abstract}
\noindent The achievement in cooling alkali atomic gases, such as
$^{87}$Rb, $^{23}$Na and $^{7}$Li, to quantum degeneracy opens up
a way to study magnetism in spinor bosons, because these
constituent atoms usually have a hyperfine spin degree of freedom.
This article reviews several basic problems related to the
ferromagnetic phase transition in spinor atomic Bose gases from a
theoretical perspective. After a brief discussion on various
possible origins of the ferromagnetic interaction, the phase
diagram of the ferromagnetically coupled spinor bosons is
investigated. It is found that the ferromagnetic transition occurs
always above Bose-Einstein condensation and the Bose condensate is
fully polarized. The low-lying collective excitations of the
spinor condensate, including spin and density modes, are
discussed. The spectrum of the density mode is of the Bogliubov
form and the spin wave spectrum has a $k^2$-formed dispersion
relation at long wavelengths. The spin-wave stiffness coefficient
contains contributions from both the ``normal" and the condensed
part of the gas.
\end{abstract}

\vspace*{-13cm}
\begin{flushleft}
{\scriptsize Chapter 6 in \textit{ 
Progress in Ferromagnetic Research} \\ \vspace*{-0.1cm} 
Editor: V.N. Murray (Nova Science Publishers, New York, 2005)}
\end{flushleft}\vspace*{13cm}

\section{Introduction}
Itinerant ferromagnetism of Fermi (electron) gases has long been a
research topic in solid state physics\cite{mohn}. This phenomenon
has already been well understood today, although not everything is
clear. Models based on a picture of almost free electrons is
fairly useful in describing the ferromagnetism. It is well-known
that a free electron gas exhibits paramagnetism (called Pauli
paramagnetism), owing to the existence of the Fermi surface. If
there is a magnetization density ${\bf M}$ in the gas, the Fermi
surfaces for spin-up and spin-down electrons are split and
consequently the band energy would be increased. When an effective
ferromagnetic (FM) exchange $I_s$ is present, electron gases can
exhibit ferromagnetism. Within the framework of the
Hartree-Fock-Stoner theory, $I_s$ results in a negative molecular
field energy when ${\bf M}$ is finite. As long as $I_s$ is large
enough, the value of the molecular field energy becomes larger
than that of the increase of the band energy caused by ${\bf M}$.
In this case, a FM ground state is energetically favored. The
critical value of $I_s$ is called the Stoner threshold.

Relatively, the magnetism of Bose gases was less studied in
history. The reason why it was so may be partially owing to the
fact that liquid $^4$He, the most prototypical Bose system being
studied in earlier years, is a system of scalar particles and does
not display magnetism at all. The first example of Bose gases with
internal degrees of freedom should be the cold atomic hydrogen
which received much attention in 1980s. The hydrogen has hyperfine
spin $F$ and thus it had ever attracted some theoretical interest
in studying its magnetic properties. However the original purpose
in surveying this atomic gas is to explore gaseous Bose-Einstein
condensation (BEC), which is the intrinsic phase transition in the
Bose gas. Since the ultimate goal of BEC was not yet attained at
that time, efforts towards other aspects, e.g. the magnetism, were
quite limited.

Great changes took place in 1995, when BEC was experimentally
realized in alkali atomic gases, such as $^{87}$Rb\cite{anderson},
$^{23}$Na\cite{davis} and later in $^{7}$Li\cite{bradley}. Since
then, research works on the physics of ultracold atomic gases have
grown explosively in the communities of atomic physics, quantum
optics and many-body physics\cite{dalfovo,leggett}. Meantime,
research interest has already gone beyond BEC itself. Alkali atoms
also have hyperfine spins, as the atomic hydrogen does. So
magnetism of spinor atomic gases is now among the most active
research topics in this field.

Earlier experiments leading to BEC in alkali atoms were performed
in magnetic traps. A relatively strong external magnetic field $H$
was applied to confine the BEC system. Because the atomic spin
direction adiabatically followed $H$, the spin degree of freedom
was frozen. As a result, the atoms behaved like scalar particles
although they carried spin. In 1998, Ketterle's group at MIT
succeeded in confining the atomic condensate in optical
traps\cite{ketterle}. The condensate was first produced in an
magnetic trap, then transferred into the optical trap for further
study. In 2001, Barrett {\it et al.} realized all-optical
formation of BEC in which the Bose condensate was created directly
in the optical trap\cite{barrett}. The optical trap is a radiation
electric field with a intensity maximum in space created by
focusing laser beams. If the frequency of light is detuned to the
red, the energy of a ground-state atom has a spatial minimum, and
then the atom is confined by the electric field. A far-detuned
optical trap can confine all the hyperfine states equally. With
the external magnetic field sufficiently low, the spin degree of
freedom of optically trapped atoms remains active. So
investigating magnetic properties of spinor condensates becomes
experimentally possible.

Theoretical investigation of ferromagnetism in spinor bosons can
be traced back to the early studies on the atomic hydrogen. Siggia
and Ruckenstein proposed that the magnetically trapped
spin-polarized hydrogen could exhibit a coherent ferromagnetism,
based on a phenomenological model\cite{siggia}. The spontaneous
magnetization is perpendicular to the stabilizing magnetic field.
Yamada considered an ideal spin-$1$ Bose gas\cite{yamada} and
argued that the spinor bosons could exhibit an intrinsic
ferromagnetism associated with BEC. He showed that the
magnetization $M(H)$ of the ideal spinor Bose gas was finite
even if $H=0$ once BEC took place, which suggested that the system
is magnetized spontaneously\cite{yamada,simkin}. He called this
phenomenon the Bose-Einstein ferromagnetism. Caramico D'Auria
{\it et al.} drew the same conclusion for interacting bosons (in
which the interaction is spin-independent)\cite{caramico}. These
results appropriately reveal that the spinor Bose gas is rather
apt to be magnetized by an external magnetic field.

The realization of optical confining of atomic condensate
significantly provoked theoretical interest in magnetism of alkali
atoms. Soon after the success of the MIT group, Ho\cite{ho} and
Ohmi and Machida\cite{ohmi} studied the spinor nature of the $F=1$
atomic condensate. They pointed out that a hyperfine spin-spin
exchange interaction coming from the s-wave scattering can be the
dominant interaction in these gases. The condensate can be either
ferromagnetic or ``polar", depending on the sign, but regardless
of the value, of the exchange interaction. They also discussed the
collective excitations in these two kinds of condensates,
including density and spin modes. Spectra of the density mode are
of the Bogliubov form in both cases, while the spin-wave spectra
have different dispersion relation.

The reason why the spinor bosons are so sensitive to the external
field and the spin-dependent interaction is attributed to the
following fact: without spin-dependent interactions, the ground
states of spinor bosons are degenerate and both the ferromagnetic
and the ``polar" states are among the ground states\cite{suto}.
Therefore an infinitesimal external field or spin-dependent
interaction can lift the degeneracy, leading to a ferromagnetic or
polar state.

In this article, some recent progress on the ferromagnetism in
spinor atomic bosons is reviewed. In Section 2 we explore the
origin of the ferromagnetism. Various mechanisms of the
ferromagnetic couplings between atomic bosons are discussed. The
phase diagram of the ferromagnetically coupled spinor bosons is
studied in Section 3. Critical temperatures of both BEC and the FM
transition are calculated. The Goldstone modes accompanying the
two phase transitions are discussed in Section 4, with special
attention to the spin wave. Our discussions are dedicated to the
3-dimensional homogeneous Bose system in the thermodynamic limit.
Experimentally, atomic gases are in a different situation. Some of
those experimental conditions are briefly discussed in the final
section.

Bosons with internal degree of freedom could be formed in Fermi
gases through spin-triplet Cooper pairing, because a pair of bound
fermions as a whole behaves like a boson. The triplet Cooper
pairing has been observed in superfluid $^3$He and in some solid
state materials. Although the Cooper pair is not identical to a
boson because Cooper pairing is not local and different pairs are
strongly overlapped in space, one can expect that triplet Cooper
pairs exhibit similar behaviors to those of the spinor bosons in
some aspects. In this article, ferromagnetism in
triplet-Cooper-paired Fermi gases is also concerned.

\section{Spin-dependent interactions in atomic bosons}
Interatomic forces in dilute atomic gases are rather weak. The
spin-dependent interactions are even much smaller than the
spin-independent one. Nevertheless, they are key ingredient in
understanding magnetism of the spinor bosons. Up to now, the
following spin-dependent interactions have been theoretically
discussed. In some cases, they can be ferromagnetic.

\subsection{Scattering between different internal states}
The interaction between atoms $V_{at}({\bf r})$ should be a
function of the separation ${\bf r}$ of the two atoms. However, it
is impossible to evaluate the ${\bf r}$-dependence of the
interaction in detail, especially at short separations. When ${\bf
r}$ is comparable to the atomic size, the standard
Born-Oppenheimer separation of the centers of mass and the
internal electronic structures of involved atoms may break down.
Consequently, atoms can not be regarded as point particles any
longer and $V_{at}({\bf r})$ is not even definable\cite{leggett}.
Fortunately, ${\bf r}$ is relatively large in dilute atomic gases,
typically of order $10^2$ nm, at which the Born-Oppenheimer
approximation should be good enough and $V_{at}({\bf r})$ could be
well defined and approximated by the van der Waals interaction. To
avoid having to calculate short range interactions in detail, the
true interatomic interaction $V_{at}({\bf r})$ may be replaced by
an effective interaction\cite{pethick},
\begin{eqnarray}
U({\bf r}={\bf r_1-r_2})=\frac {4\pi \hbar^2 a}{m} \delta({\bf r_1-r_2}),
\end{eqnarray}
where $a$ is the low-energy s-wave scattering length. The s-wave
scattering perfectly describes the interactions of dilute alkali
atomic gases, with $a$ is of the order of $100 a_B$ where $a_B$ is
the Bohr radius.

In case that involved atoms have internal degree of freedom, for
example the hyperfine spins for alkali atoms, interatomic
interactions may give rise to transitions between different
sub-states. Two atoms in the initial state $|\alpha,\beta\rangle$
may be scattered by collisions to the state
$|\alpha^\prime,\beta^\prime\rangle$, where $\alpha$ and $\beta$
denote the hyperfine states of the two atoms. In this case the
scattering is a multi-channel problem and the scattering length
could vary according to channels. That is to say, the scattering
length $a_{\alpha\beta,\alpha^\prime\beta^\prime}$ is
''spin-dependent", with a large number of free parameters. The
number of free parameters could be reduced by considering particle
exchange, time-reversal and especially spin rotational symmetries.
Suppose that the collision between atoms does not change the
hyperfine spin $F$ (this is usually the case for sufficiently low
energy collisions if the atoms are in the lowest hyperfine
multiplet) of the individual atoms, the pairwise interaction keeps
rotationally invariant in the hyperfine spin space. Then the
collisions between two atoms with hyperfine spin $F_1$ and $F_2$
only depend on the total spin $f=F_1+F_2$ and $U({\bf r_1-r_2})$
is reduced, in the s-wave limit, to\cite{ho}
\begin{eqnarray}\label{u1}
U({\bf r_1-r_2}) = \frac {4\pi \hbar^2 }{m} \delta({\bf r_1-r_2})
\sum_f a_f {\mathcal{P}}_f,
\end{eqnarray}
where $a_f$ is the scattering length for collisions between two
atoms with total spin $f$, and $\mathcal{P}_f$ is the projection
operator which projects the total spin of the pair of atoms into
$f$ channel. $\mathcal{P}_f$ satisfies the condition $\sum_f
\mathcal{P}_f=1$. For two bosons with spin $F$, $f$ takes the
values $f=2F, 2F-2, 2F-4,...,0$. The $f=2F-1, 2F-3,...$ channels
are forbidden owing to the exchange symmetry of two bosons.
Likewise,
\begin{eqnarray}\label{u2}
{\bf F}_1\cdot {\bf F}_2 = \sum_f \lambda_f \mathcal{P}_f ,
\end{eqnarray}
with $\lambda_f=[f(f+1)-2F(F+1)]/2$. Combining Eqs. (\ref{u1}) and
(\ref{u2}), one can get the effective interaction in term of the
hyperfine spin operators. In particular, the effective interaction
for the spin-$1$ bosons is given by\cite{ho,ohmi}
\begin{eqnarray}\label{eq3}
U({\bf r_1-r_2}) = \frac {4\pi \hbar^2 }{m} \left [ g_0  +
   g_2 {\bf F}({\bf r_1})\cdot {\bf F}({\bf r_2 }) \right]
   \delta({\bf r_1-r_2}) ,
\end{eqnarray}
where $g_0 = ({a_0+2a_2})/3$, $g_2=({a_2-a_0})/3$, and ${\bf
F}=(F^x,F^y,F^z)$ are $3\times 3$ spin matrices. The $g_2$ term is
a Heisenberg-like exchange interaction. Usually the difference
between $a_2$ and $ a_0$ is very small, so $g_2$ is much smaller
than $g_0$.

Why could scattering lengths be spin-dependent? To answer this
question, we recall that atoms have internal electronic
structures. The pairwise interaction of two alkali atoms depends
on the spin configuration of the two valence electrons. For
example, the interaction has an attractive contribution when two
valence electrons are in the singlet state since they can occupy
the same orbital. Otherwise, this contribution is absent.

The exchange interaction can be ferro- or antiferro-magnetic,
depending on $a_0$ is greater or less than $a_2$. In principle,
$a_0$ and $a_2$ can be calculated directly by solving
Schr\"{o}dinger equation. But exact solutions are not get-at-able
for many electron atoms. It is possible to check the sign of $g_2$
indirectly by studying spin dynamics of the spinor Bose
condensate, since behaviors of the ferro- and antiferro-magnetic
spinor condensates are quite different\cite{ketterle,schma,chang}.
Both theory and experiment suggested that $g_2<0$ for the gas of
$F=1$ $^{87}$Rb \cite{ketterle,burke}, and $g_2>0$ for $F=1$
$^{23}$Na \cite{ketterle,crubellier}. The spin exchange is
antiferromagnetic for the $F=2$ $^{85}$Rb and $^{87}$Rb
atoms\cite{schma,chang,burke}.

\subsection{Super-exchange process}
Now let us consider a Bose gas moving on a periodic optical
lattice. Such a system can be approximately described by the boson
Hubbard model\cite{greiner,fisher},
\begin{eqnarray}
H_{Hubbard} = - \sum_{\langle ij \rangle} t_{ij} a^{\dag}_{i} a_{j}
    + \frac 12 U_0 \sum_{i} n_{i} (n_{i}-1) .
\end{eqnarray}
Here ${\langle ij \rangle}$ labels two nearest neighbor sites on
the lattice, $U_0$ is the on-site Hubbard repulsion. The optical
lattice is produced by the interference of laser beams. Two
counter-propagating laser beams form a standing wave, which acts
on the atoms as a periodic potential, $V({\bf r})=\sum_i V_i {\rm
sin}^2k_i r_i$, where $k_i$ is the wave vector of the laser. For a
given optical potential, $t_{\langle ij \rangle}$ and $U_0$ are
readily evaluated\cite{jaksch}. The Hubbard repulsion is
proportional to the s-wave scattering length.

As is well-known, the electron Hubbard model is used to describe
the Mott insulator in condensed matter physics. Provided the
Hubbard repulsion is large enough in comparison to the
hopping matrix $t_{ij}$, the energy band of electrons splits into
two branches separated by a energy gap. At half filling, the lower
Hubbard band is fully filled while the upper one is empty and thus
electrons are in the Mott insulating state. With the charge degree
of freedom being frozen, the low-energy behaviors of electrons can
be described by an effective spin model. One can derive the
effective spin Hamiltonian directly from the Hubbard model through
perturbation approach, taking the hopping term as the
perturbation. To the second order of the perturbation, the spin
Hamiltonian is described by the Heisenberg model. The spin coupling
between localized electrons, called the super-exchange,
is {\it antiferromagnetic}.

The boson Hubbard model is introduced to account for the
superfluid-Mott insulator transition in lattice
bosons\cite{fisher,jaksch}. It is argued that bosons are
insulating in the large $U_0$ limit, as the correlated electrons
do. Following the suggestion of \protect\cite{jaksch}, the
superfluid-Mott insulator transition has already observed
experimentally\cite{greiner}. By analogy, one can expect that the
{\it spinor} boson Hubbard model is reduced to an effective spin
Hamiltonian, too.

Let us first look at a two-component Bose system, which can be
mapped into a pseudo-spin-$\frac 12$ boson Hubbard model. Yang
{\it et al.}\cite{yang} and Duan {\it et al.}\cite{duan} have
derived the effective spin Hamiltonian,
\begin{eqnarray}
H = J \sum_{{\langle ij \rangle}} {\bf S_{i}}\cdot {\bf S_{j}} ,
\end{eqnarray}
which is just the Heisenberg model. However, the super-exchange
interaction between localized bosons is {\it ferromagnetic},
$J =-{4t^2}/U_0$, as opposite to electrons.

For bosons with integer spins, the effective spin Hamiltonian is
relatively complicated. Imambekov {\it et al.} has investigated
the spin-1 boson Hubbard mode in detail\cite{imambekov}. To order
of ${t^2}/U_0$, the spin Hamiltonian consists of two terms,
\begin{eqnarray}
H = J_1 \sum_{{\langle ij \rangle}} {\bf S_{i}}\cdot {\bf S_{j}}
    + J_2 \sum_{{\langle ij \rangle}} ({\bf S_{i}}\cdot {\bf S_{j}})^2 ,
\end{eqnarray}
with $J_1=J_2=-{2t^2}/U_0$. The first term tends to stabilize a
ferromagnetic order, while the second term favors a local singlet
of the two coupled spins.

In above derivations we assume that the number of bosons is the
same as the number of lattice sites, thus each site is occupied by
only one particle. But for bosons, any occupation is allowed. The
system could be a Mott insulator as long as the boson density is
commensurate. {Ref. \protect\cite{imambekov}} presented detailed
discussions on the spin exchange at various commensurate
occupations. Moreover, an antiferromagnetic on-site coupling
between bosons is considered in {Ref. \protect\cite{imambekov}}.
$J_2/J_1$ increases with increasing either the on-site
antiferromagnetic coupling or the occupation number, and the
system tends to have a nematic rather than a ferromagnetic ground
state.

\subsection{Magnetic dipolar interaction}
Corresponding to the (hyperfine) spin degrees of freedom, spinor
atomic bosons have a magnetic moment ${\bf m}=\gamma {\bf S}$,
which induces the magnetic dipolar interaction between particles,
\begin{eqnarray}
U_{md} = \frac {\mu_0}{4\pi r^3}[{\bf m}_1\cdot{\bf m}_2
        - 3({\bf m}_1\cdot {\bf r})({\bf m}_2\cdot {\bf r})] ,
\end{eqnarray}
where $\mu_0$ is the vacuum permeability. The influence of the
magnetic dipolar interaction on the properties of spinor
Bose-Einstein condensates has recently attracted considerable
interest\cite{pfau,pu,gross,Gu1}. Although very weak, it is
expected to largely enrich the variety of phenomena in ultracold
Bose gases, especially in the gases of atoms with larger magnetic
moments, such as europium where $m=7\mu_B$ ($\mu_B$ is the Bohr
magneton) and Chromium where $m=6\mu_B$\cite{kim}.

The magnetic dipolar interaction is anisotropic and its strength
depends not only on the separation but also on the spin
configurations of the two interacting particles. Pu {\it et al.}
studied the magnetism of an ensemble of spinor condensates
confined in a lattice\cite{pu,gross}. The ``minicondensate" at
each site behaves as a {\it localized} mesoscropic magnet and
interacts with each other via the magnetic dipolar interaction.
They showed that the ground state is ferromagnetic in a chain, and
antiferromagnetic in a square lattice.

However it is rather difficult to treat precisely the dipolar
interaction between particles in a {\it gas}, because of the
anisotropic feature and the long-range character of the
interaction. Many efforts have been done to study the ground state
of the magnetic (and electric) dipolar gases (or fluids). It was
reported that a ferromagnetic/ferroelectric nematic order is
favored under certain conditions\cite{wei,osipov}. Recently, the
dipolar ferromagnetism has also been predicted in ensembles of
randomly distributed nano-particles\cite{denisov}.

Let's demonstrate why the magnetic dipolar interaction can result
in ferromagnetism from a mean-field viewpoint\cite{Gu1}. When all
the magnetic moments are arranged parallel, an effective magnetic
field ${\bf B}=\mu_0 {\bf M}$ is created inside the polarized body
where ${\bf M}$ is the magnetization density, owing to the
superposition of the intrinsic field of all the magnetic moments.
The energy density of the effective magnetic field is $f_m =
{B^2}/{(2\mu_0)}={\mu_0} M^2/2$. On the other hand, the magnetic
moment of spinor bosons does respond to the internal magnetic
field, so the spin direction should follow ${\bf B}$, which leads
to an energy decrease: $f_c=-{\bf M\cdot B}=-\mu_0 M^2$. Therefore
the total free energy density is negative, $f_m+f_c = -{\bf \mu_0}
M^2/2$. Contrarily, in the ``polar" or ``equal spin" states the
magnetic moments of particles are compensated: $M=0$. At the
mean-field level, an internal reference particle can not sense the
moments of other particles and the magnetic free energy is zero.

It is worth noting that the dipolar ferromagnetism could manifest
itself more easily in a Bose gas than in a Boltzmann gas or a
Fermi gas\cite{Gu1}. The ferromagnetically ordered state can be
destroyed due to the entropy increase at finite temperatures. So a
Boltzmann gas can show the dipolar ferromagnetism only at low
temperatures comparable to the energy scale of the dipolar
interaction. Since the Bose condensate has no entropy, the
ferromagnetic state could survive in the Bose condensed particles
below the BEC temperature which is relatively high in comparison
to the energy scale of the dipolar interaction. As for the Fermi
gas, it can hardly display the dipolar ferromagnetism, because the
dipolar interaction is too weak to reach the Stoner threshold for
the itinerant ferromagnetism. That's why the dipolar interaction
plays a less important role than exchange interactions in
understanding magnetism in solid. The latter is much stronger.

\section{Phase diagram of ferromagnetic spinor Bose gases}
As suggested in the Weiss molecular field theory (for classical
systems) and the Stoner theory (for Fermi systems)\cite{mohn}, the
ferromagnetic coupling is expected to induce a ferromagnetic
transition in Bose gases. On the other side, in a Bose system
whose particle number is conserved there exists an intrinsic phase
transition, Bose-Einstein condensation. It is natural to suppose
that the FM transition temperature $T_F$ depends on the energy
scale of the coupling $I$. Therefore $T_F$ is possibly smaller
than $T_c$, the BEC temperature, for small values of $I$, and
$T_F>T_c$ if $I$ is sufficiently large. Is that true?

In this section we study how the FM transition and BEC emerge in a
spinor Bose system. We first calculate the phase transitions by
dealing with a simple microscopic model, then analyze the
predicted phase diagram from a phenomenological point of view. For
simplicity, we consider a 3D homogeneous spinor Bose gas.

\subsection{A microscopic model}
In {Ref.\ \protect\cite{Gu2}} we have investigated the interplay
between the FM transition and BEC. The calculation starts with a
simplified Hamiltonian given by
\begin{eqnarray}
H = - t \sum_{{}\sigma}{a}^{\dag}_{i\sigma} {a}_{j\sigma}
    - I_H \sum_{{\langle ij \rangle}}{\bf S}_i\cdot {\bf S}_j .
\end{eqnarray}
The particles are treated as being on some kind of ``lattice".
$\langle ij \rangle$ denotes two nearest-neighboring sites. The
first term in the Hamiltonian represents the kinetic energy. In
following calculations it is replaced by the kinetic energy of
free particles with mass $m^*$, $\epsilon_{\bf
k}=\hbar^2k^2/2m^*$.

The second term describes the FM coupling, which could be
decoupled via the mean-field approximation,
\begin{eqnarray}
-\sum_{\langle ij \rangle}{\bf S}_i\cdot {\bf S}_j \approx
     - \sum_{\langle ij \rangle} ( \langle{\bf S}_i\rangle \cdot {\bf S}_j
     + {\bf S}_i \cdot \langle{\bf S}_j\rangle
     -\langle{\bf S}_i\rangle \cdot \langle{\bf S}_j\rangle ),
\end{eqnarray}
where ${\bf M}_i=\langle{\bf S}_i\rangle$ serves as the
ferromagnetic order parameter, and is chosen to be along the $z$
direction, $\langle{\bf S}_i\rangle=(0,0,M_i)$ and ${M}_i =
\langle{S_i^z}\rangle = \sum_{\sigma}\sigma \langle
a_{i\sigma}^\dag a_{i\sigma} \rangle$ where $a_{i\sigma}$
annihilates a boson with spin quantum number $\sigma$ at site $i$.
We take $M_i=M$ for a homogeneous boson gas. Following the Stoner
theory for fermion gases, we call $H_m = Z I_H M$ the molecular
field and $I_s=Z I_H$ the Stoner exchange, where $Z$ is the
effective ``coordination number" which is an irrelevant
dimensionless parameter of order unity for a gas. Then the
effective Hamiltonian reads
\begin{eqnarray}
H-N\mu= \sum_{{\bf k}\sigma}
   ( \epsilon_{\bf k} - \mu - \sigma H_m ) n_{{\bf k}\sigma} ,
\end{eqnarray}
where $\mu$ is the grand canonical chemical potential.

The mean-field self-consistent equations consist of two equations,
$$n = \frac 1V\sum_{{\bf k}\sigma}
      \langle n_{{\bf k}\sigma} \rangle, ~~
  M = \frac 1V\sum_{{\bf k}\sigma}
      \sigma \langle n_{{\bf k}\sigma} \rangle.$$
$n=N/V$ is the density of particles, $V$ is the volume of the
system and $N$ is the total number of particles. $M$ is the
magnetization density. Let us consider a $F=1$ Bose gas with the
self-consistent equations given by
\begin{subequations}
\begin{eqnarray}\label{eq11a}
1 &=& {\overline n}_0 + t^{\frac 32} \left[
     f_{\frac 32}(a) + f_{\frac 32}(a+b) + f_{\frac 32}(a+2b)  \right] ,
     \\ \label{eq11b}
{\overline M} &=& {\overline n}_0 +
   t^{\frac 32}\left[ f_{\frac 32}(a)  - f_{\frac 32}(a+2b) \right] ,
\end{eqnarray}
\end{subequations}
where $a=-(\mu+H_m)/(k_B T)$, $b=H_m/(k_B T)$, the reduced
temperature and coupling are given by $t={k_B T m^*}/{(
2\pi\hbar^2{\overline n}^{2/3} )}$ and $I={I_s n^{1/3} m^*}/{(
2\pi\hbar^2 )}$, respectively, $n_0$ is the condensate density,
${\overline n}_0=n_0/n$ is the condensate fraction, ${\overline
M}=M/n$ is the normalized magnetization, and $f_{s}(a)$ is the
polylogarithm function defined as
\begin{eqnarray}
f_{s}(a) \equiv {\rm Li}_s( e^{-a})
         = \sum_{p=1}^{\infty}\frac {\left(e^{-a}\right)^p}{p^s} .
\end{eqnarray}
We note that $f_s(0)=\zeta(s)$, the Riemann zeta function.

Based on this simple model, we have shown that an {\it
infinitesimal} FM coupling can induce a FM phase transition at a
finite temperature above BEC. As is well-known, the BEC critical
temperature $t_c$ can be determined by calculating the chemical
potential\cite{huang}: $a>0$ for $t>t_c$ and $a\to 0$ as $t\to
t_c$ from above. Assume a FM phase transition is induced by $I$ at
the transition temperature $t_F$. We first suppose $I$ is very
large so that $t_F>t_c$. Provided that the FM transition is {\it
continuous}, i.e., $b\to 0$ with $t \to t_F$, the mean-field
equations become
\begin{subequations}
\begin{eqnarray}
1 &= 3 t_F^{\frac 32} f_{\frac 32}(a_F) ,\\
1 &= 2 I t_F^{\frac 12} f_{\frac 12}(a_F) .
\end{eqnarray}
\end{subequations}
where $a_F=a(t_F)$. These equations define a relation between
$t_F$ and $I$. $I$ is a monotically decreasing function of $a_F$.
As $a_F\to 0$, $f_{\frac 32}(a_F)\to \zeta(3/2)\approx 2.612$, and
$f_{\frac 12}(a_F)\approx \sqrt{\pi/a_F}$. So for small values of
$I$ and $a_F$ we have
\begin{eqnarray}
a_F \approx {4\pi} t_0 I^2 ,
\end{eqnarray}
where $t_0=1/[3\zeta(3/2)]^{2/3}$ is the reduced BEC critical
temperature for the Bose gas with $I=0$. $a_F$ is larger than zero
at $t_F$, which means BEC does not yet take place. The Bose gas
undergoes BEC at a lower temperature.

\begin{figure}
\center{\epsfxsize=70mm \epsfysize=60mm \epsfbox{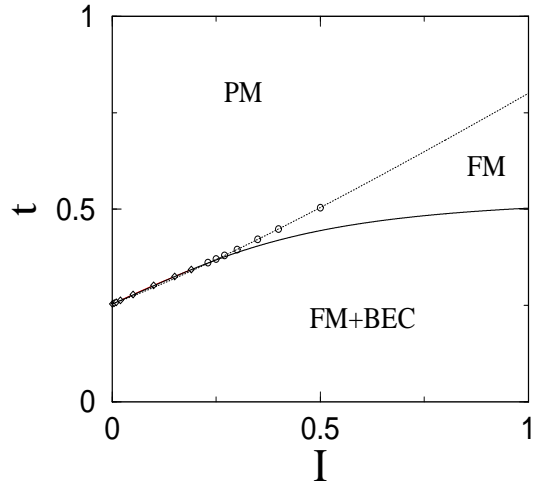}}
\caption{ \label{fig:epsart} Reduced temperature $t$ vs. reduced
FM coupling $I$ phase diagram of spin-1 Bose gases. The
paramagnetic normal phase (PM), ferromagnetic normal phase (FM),
and coexisting phase of ferromagnetism and Bose condensation
(FM+BEC) are indicated. }
\end{figure}

By solving the mean-field equations numerically, we can obtain
that both the FM transition temperature and the BEC critical
temperature increase with the FM coupling. Asymptotic analysis at
very small $I$ shows that
\begin{eqnarray}
\frac {\delta t_F}{t_0} \sim \frac {\delta t_c}{t_0} \propto I ,
\end{eqnarray}
where $\delta t_F=t_F-t_0$ and $\delta t_c=t_c-t_0$. We should
note that the asymptotic analysis is not appropriate for this
simple model within mean-field approximation, because numerical
results show that both FM transition and BEC are discontinuous at
very small $I$. Approximately, the critical value of $I$ under
which the transition becomes discontinuous is $0.35$ for the FM
transition and $0.2$ for BEC. For larger couplings, the FM
transition is well described as being continuous. This point is
consistent with the Weiss molecular field theory for classical
particles, in which the FM transition is continuous\cite{mohn},
because for large couplings, the FM transition occurs at a
relatively high temperature, when the Bose statistics reduces to
Boltzmann statistics.

Figure 1 shows the phase diagram of spin-1 Bose gases. The curves
are obtained by solving the reduced self-consistent equations
supposing both FM transition and BEC are continuous. Diamonds and
circles denote numerical results of Eqs. (\ref{eq11a}) and
(\ref{eq11b}). In order to make a comparison with the Fermi gas,
Fig. 2 plots the schematic phase diagram for both the Bose and
Fermi gases together.

\begin{figure}
\center{\epsfxsize=75mm \epsfysize=60mm \epsfbox{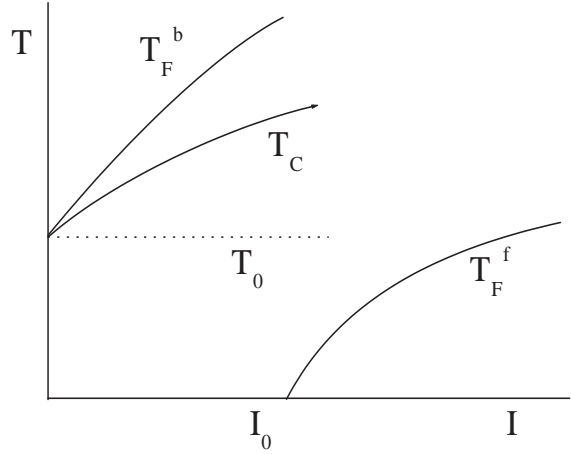}}
\caption{ \label{fig:epsart} Schematic relations between
transition temperatures and FM couplings $I_s$. $T^b_F$ and
$T^f_F$ represent the FM transition temperature for Bose and Fermi
gases respectively. $T_c$ and $T_0$ denote the BEC critical
temperature for spinor bosons with and without couplings
respectively. $I_0$ is the Stoner threshold. }
\end{figure}

\subsection{Phenomenological analysis}
In this subsection, we examine the above-predicted phase diagram
on the basis of Ginzburg-Landau (GL) phenomenological
theory\cite{Gu1}. The advantage of the GL theory is that one can
describe properties of a system by analyzing the symmetry of the
system in a simple way, without a detailed knowledge of the
microscopic background. It is applicable to the case of
interacting bosons, in which the phase transitions, both FM
transition and BEC, are supposed to be of second order.

To begin with, we derive an appropriate GL free energy density
functional that describes the coexistence of BEC and
ferromagnetism. Generally, such a free energy density consists of
three different parts\cite{Gu1,Gu3}:
\begin{eqnarray}
f_t({\bf \Psi},{\bf \Psi}^\dag,{\bf M}) = f_b({\bf \Psi},{\bf
\Psi}^\dag) + f_m({\bf M}) + f_c({\bf \Psi},{\bf \Psi}^\dag,{\bf
M}),
\end{eqnarray}
corresponding to the Bose condensed phase, the ferromagnetic phase
of the normal gas and the coupling between the two phases. Here
${\bf \Psi}^{\dag}\equiv ({\bf \Psi}^T)^{*}=(\Psi_F^{*},
\Psi_{F-1}^{*},..., \Psi_{-F}^{*})$ is the complex order parameter
of the spinor Bose condensate. The condensed Bose gas is described
by the two-fluid model. We suppose $f_m({\bf M})$ only represents
the ferromagnetic phase of the normal gas, with the order
parameter ${\bf M}$ proportional to the magnetization density.

Following Ginzburg and Pitaevskii\cite{ginzburg}, the free energy
density of an {\it isotropic} spin-$F$ Bose-Einstein condensate is
modelled as
\begin{eqnarray}\label{eq17}
f_b = \frac {\hbar^2}{2m} \nabla {\bf \Psi}^{\dag} \cdot \nabla {\bf \Psi}
      + \alpha |{\bf \Psi}|^2 + \frac {\beta_0}2 |{\bf \Psi}|^4  +
      \frac {\beta_s}2 \Psi^*_\sigma \Psi^*_{\sigma'}{\bf F}_{\sigma\gamma} \cdot
      {\bf F}_{\sigma'\gamma'} \Psi_{\gamma'} \Psi_{\gamma}  ~,
\end{eqnarray}
where $|{\bf \Psi}|^2={\bf \Psi}^{\dag}{\bf \Psi}$ and repeated
sub-indices represent summation taken over all the hyperfine
states. $\alpha=\alpha^\prime(T-T_c^0)$ and $T_c^0$ is the BEC
critical temperature. Both $\alpha^\prime$ and $\beta_0$ are
positive parameters, and $\beta_0$ contains contributions of the
spin-independent interactions. The fourth term has SO(3) symmetry,
arising from the spin-exchange interactions. $\beta_s$ can be
positive or negative, depending on the exchange interaction
antiferromagnetic or ferromagnetic. Due to the $\beta_s$ term, the
time-reversal symmetry in the Bose condensate should be broken
spontaneously.\footnote{Using the notation $(\Psi_1, \Psi_0,
\Psi_{-1})=(\phi_1 e^{i \theta_1}, \phi_0, \phi_{-1} e^{i
\theta_{-1}})$, the $\beta_{s}$ term becomes ${\beta_{s}} \phi^2_0
[ \phi^2_1 + \phi^2_{-1} +2 \phi_1\phi_{-1}
\cos(\theta_1+\theta_{-1}) ] + \beta_{s} (\phi^2_1 -
\phi^2_{-1})^2/2$. Since $\theta_1+\theta_{-1}=\pi$ for
$\beta_{s}>0$ and $\theta_1+\theta_{-1}=0$ for $\beta_{s}<0$, it
therefore determines the ground state of the condensate: it is
ferromagnetic for $\beta_{s}<0$ and ``polar" for
$\beta_{s}>0$\cite{ho, ohmi}.} Hereinafter we consider the
ferromagnetic case ($\beta_{s}<0$).

As mentioned in the last subsection, a FM transition in normal gas
happens above BEC. The free energy density for this ferromagnetic
phase can be expanded in powers of $|{\bf M}|^2$:
\begin{equation}\label{eq18}
f_m = c |\nabla {\bf M}|^2
    + a^\prime(T-T_f)\frac {|{\bf M}|^2}2 + b \frac {|{\bf M}|^4}4 ,
\end{equation}
where $a=a^\prime(T-T_f)$, $c$, $a^\prime$ and $b$ are positive
constants, $T_f$ is the FM transition temperature in the normal
Bose gas. We suppose that the ferromagnetic normal gas couples
linearly to the spinor condensate,
\begin{equation}
f_c = -g {\bf M} \cdot \Psi^*_{\sigma}{\bf F}_{\sigma\gamma}
      \Psi_{\gamma} ~,
\end{equation}
with the coupling constant $g>0$. This is the simplest coupling
term that satisfies the physical situation.

We do not consider fluctuations in this subsection, so all the
parameters are supposed to be given by their average: $\langle
\Psi_\sigma \rangle = \Phi_\sigma $ and $\langle {\bf M} \rangle
=(0,0,M_0)$ where the FM order parameter is chosen to be along the
$z$ direction for convenience. $\Phi_\sigma^{\dag}\Phi_\sigma=n_0$
is just the density of condensed bosons. Hence the gradient terms
in Eqs. (\ref{eq17}) and (\ref{eq18}) can be dropped for the
homogeneous system.

Minimizing the total free energy $f_t$ with respect to
$\Phi^\dag$, one gets
\begin{eqnarray}
[\alpha^\prime(T-T_c^0) - gM_0\sigma]\Phi_\sigma + \beta_0 |{\bf
\Phi}|^2 \Phi_\sigma
   + \beta_s \Phi^*_{\sigma'}{\bf F}_{\sigma\gamma} \cdot
     {\bf F}_{\sigma'\gamma'} \Phi_{\gamma'} \Phi_{\gamma} = 0~.
\end{eqnarray}
Then the stable solution reads
\begin{eqnarray}
|\Phi_0|^2 = |\Phi_{-1}|^2 = 0,~~~|\Phi_1|^2 = \frac
{\alpha^\prime}{\beta_0+\beta_s} \left( T-T_c^0-\frac
g{\alpha^\prime}M_0 \right) .
\end{eqnarray}
Only the spin-$1$ bosons condense, occurring at an enhanced BEC
transition temperature,
\begin{eqnarray}\label{eq22}
T_c=T_c^0+\frac{g}{\alpha^\prime}M_0 .
\end{eqnarray}
Obviously, the magnetization in the normal gas promotes the BEC
critical temperature. At $T=T_c$, the order parameter of the
condensate is zero, and we can derive the value of $M_0$ by
minimizing $f_m(M_0)$ with respect to $M_0$,
\begin{eqnarray}\label{eq23}
M_0 = \sqrt{\frac {a^\prime}{b}[T_f-T_c]}.
\end{eqnarray}

We now derive the phenomenological relations between the two
transition temperatures. At very small ferromagnetic couplings,
$I\to 0$, both $\delta T_c=T_c-T_c^0$ and $\delta T_f=T_f-T_c^0$
tend to zero. Substituting Eq. (\ref{eq23}) into (\ref{eq22}), one
finds
\begin{eqnarray}
\delta T_c=\sqrt{ T^* (\delta T_f-\delta T_c)} ,\nonumber
\end{eqnarray}
with $T^*=(g/\alpha^\prime)^2a^\prime/b$. Under the condition that
$\delta T_f<<T^*$ we have
\begin{eqnarray}\label{eq24}
(\delta T_c)\approx \delta T_f\left(1-\frac {\delta T_f}{T^*}\right) .
\end{eqnarray}
Suppose $\delta T_f$ increases linearly with the ferromagnetic
coupling $I$, $\delta T_f=CI$, when $I<<1$, the coupling
dependence of $\delta T_c$ is given by the formula
\begin{eqnarray}\label{eq25}
\delta T_c=CI\left(1-\frac {CI}{T^*}\right) .
\end{eqnarray}
Besides a linear term, $\delta T_c$ also depends on $I$
quadratically. So far, the phase diagram predicted in the previous
subsection is roughly reproduced. It is noteworthy that the linear
dependence of $\delta T_f$ on $I$ is only an assumption in
phenomenological theory. To decide the precise relation between
$\delta T_f$ and $I$ is definitely of theoretical interest and is
still an open question.

In analogy to the ferromagnetically coupled spinor bosons, a
similar system has been found in condensed matters, say, the
ferromagnetic superconductors\cite{saxena,pfleiderer}. In these
materials, such as $\rm UGe_2$\cite{saxena} and $\rm
ZrZn_2$\cite{pfleiderer}, coexistence of ferromagnetism and
superconductivity has been observed. The phase diagram, see Fig.
3, indicates that these materials first undergo a FM transition,
then go into the superconducting state. Considering the strong
magnetization in these materials, it is suggested that the Cooper
pairing should be $p$-wave triplet, thus behaves like spin-1
bosons in some sense. Phenomenological theory has been performed
to explain the phase diagram of FM
superconductors\cite{walker,shopova}. This subsection is just an
extension of the phenomenological theory for FM superconductors to
spinor bosons.

\begin{figure}
\center{\epsfxsize=75mm \epsfysize=60mm \epsfbox{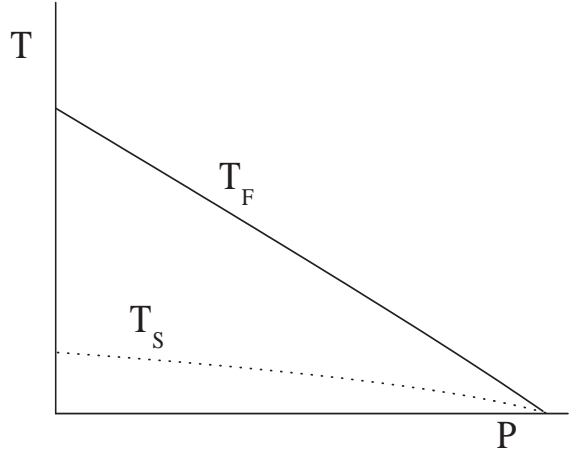}}
\caption{ \label{fig:epsart} Schematic Phase diagram of the FM
superconductor $\rm ZrZn_2$. $T_F$ and $T_{S}$ are the FM and
superconducting transition temperatures. $P$ is the pressure. }
\end{figure}

\section{Spin waves in spinor Bose condensates}
Once a continuous symmetry is spontaneously broken, gapless
Goldstone modes are expected. Spontaneous magnetization breaks the
spin rotational symmetry and the corresponding Goldstone modes are
referred as spin waves. In a ferromagnet, the dispersion relation
of the spin wave takes the form $\omega_s=c_s k^2$ where $k$ is
the wave number, and the spin-wave stiffness coefficient $c_s$
should depend on the ferromagnetic couplings. For example, the
long wave-length spectrum of spin waves in a Heisenberg
ferromagnet is $\hbar \omega \sim |J|k^2$ with the simple cubic
structure, where $J$ is the Heisenberg exchange.

The $k^2$-formed spectrum can be derived qualitatively from the GL
free energy density functional for FM transitions given by Eq.
(\ref{eq18}). Within GL theory, the Goldstone mode is interpreted
as transverse fluctuations on the average value of the order
parameter. The full order parameter of the FM phase takes the form
${\bf M} =\langle {\bf M} \rangle+\delta{\bf M} =(\delta
M_x,\delta M_y,M_0+\delta M_z)$ where $\delta M_z$ is the
longitudinal fluctuation and $\delta M_x$, $\delta M_y$ are
transverse components. To linear order in $\delta{\bf M}$, the
free energy is is given by $f_m=f^0_m+f^I_m$, with
\begin{subequations}
\begin{eqnarray}
f^0_m &=& \frac 12 a M_0^2 + \frac 14 b M_0^4 ,\\
f^I_m &=& c \nabla {\delta M_z} \nabla {\delta M_z}
      + \frac 12 \left(a+3bM_0^2\right) {\delta M_z}{\delta M_z}
      \nonumber \\
      &&+ c \nabla {\delta M_+} \nabla {\delta M_-}
      + \frac 12 \left(a+bM_0^2\right) {\delta M_+}{\delta M_-} .
\end{eqnarray}
\end{subequations}
Here $\delta M_+=\delta M_x+i\delta M_y$ and $\delta M_+=\delta
M_x-i\delta M_y$. Below FM transition point, $\partial
f^0_m/\partial M_0=0$ and we have $a+b M_0^2=0$. Under this
condition, the dispersion relation of the transverse mode becomes
gapless,
\begin{equation}
\hbar \omega_{\pm} = c k^2 ,
\end{equation}
while the longitudinal mode is gapped, $\hbar \omega_{z} = c k^2 +
bM_0^2$ and thus this mode can be neglect at low energy. Here
$c_s=c$ is a phenomenological parameter.

To proceed, we consider Goldstone modes in a spinor Bose
condensate. This problem has received much attention recently
\cite{ho,ohmi,ueda,szepfalusy,szirmai,ohtsuka}. Since both the
conservation of particle numbers and the spin rotational symmetry
are spontaneously broken in the ferromagnetic condensate, an
interesting questions arise: how the spin wave manifests itself
therein. Ho\cite{ho} and Ohmi and Machida\cite{ohmi} studied the
zero-temperature collective excitations in spinor condensates
based on the Bogliubov approximation using an equation of motion
approach. As they pointed out, the density, spin and
``quadrupolar" spin fluctuations are related to $\delta\Psi_1$,
$\delta\Psi_0$, and $\delta\Psi_{-1}$, respectively, since $\delta
n=\sqrt{n_0}(\delta\Psi_1+\delta\Psi^*_1)$, $\delta M_-
=\sqrt{n_0}\delta\Psi^*_0$ and $\delta M^2_-
=2\sqrt{n_0}\delta\Psi^*_{-1}$. A Bogliubov spectrum for density
fluctuations and a $k^2$-formed dispersion for spin fluctuations
were derived,
\begin{equation}
\hbar\omega_1 = \sqrt{\epsilon_k^2 +
      2\frac {4\pi \hbar^2 }{m}(g_0+g_s)n_0\epsilon_k},~~~~
\hbar\omega_0 = \epsilon_k = \frac {\hbar^2k^2}{2m} ,
\end{equation}
consistent with available theories of BEC and ferromagnetism. Ueda
obtained the same results through a many-body mean-field
approach\cite{ueda}. These results hold for a dilute
weak-interacting atomic gas, in which the condensate fraction is
almost equal to one at sufficiently low temperatures.

In the following, we generalize the equation of motion approach to
finite temperature cases, especially near the BEC point, making
use of GL theory\cite{Gu3}. The full order parameters for a
condensate are written as ${\bf \Psi}^{\dag} = (\Phi_1^{*}+\delta
\Psi_1^{*}, \delta \Psi_0^{*}, \delta \Psi_{-1}^{*})$. Adopting
the results of Section 3.2 that only $F=1$ bosons condense, the
free energy density for a ferromagnetic condensate reads
\begin{subequations}
\begin{eqnarray}
f_b &=& f^0_b + f^I_b,\\
f^0_b &=& \alpha {\Phi_1}^2 + \frac 12 (\beta_0+\beta_s)
    {\Phi_1}^4  ~,\\
f^I_b &=& \frac {\hbar^2}{2m} \nabla{\delta\Psi^*_\sigma}
    \nabla {\delta\Psi_\sigma} + [\alpha+ (\beta_0+\beta_s)
    {\Phi_1}^2] {\delta\Psi^*_\sigma} {\delta\Psi_\sigma}
\nonumber\\
    &&+ \frac 12 (\beta_0+\beta_s) {\Phi_1}^2 ({\delta\Psi^*_1}
     + {\delta\Psi_1})^2 -2{\beta_s} {\Phi_1}^2 {\delta\Psi^*_{-1}}
    {\delta\Psi_{-1}} ~.
\end{eqnarray}
\end{subequations}
Here $\beta=\beta_0+\beta_s$. The equations of motion are obtained
from $f^I_b$ according to $i\hbar\partial_t \delta\Psi_\sigma =
(\partial f^I_b)/(\partial\delta\Psi^*_\sigma)$. We have
\begin{equation}
i\hbar \partial_t \begin{pmatrix}\delta\Psi_1\\ -\delta\Psi^*_1 \end{pmatrix}
  = (\epsilon_k + \alpha+ \beta{\Phi_1}^2)
    \begin{pmatrix}\delta\Psi_1\\ \delta\Psi^*_1 \end{pmatrix}
  + \beta{\Phi_1}^2 \begin{pmatrix}\delta\Psi_1+\delta\Psi^*_1\\
    \delta\Psi_1+\delta\Psi^*_1 \end{pmatrix}
\end{equation}
\begin{equation}
i\hbar \partial_t \begin{pmatrix}\delta\Psi_0\\ \delta\Psi_{-1} \end{pmatrix}
  = \begin{pmatrix}(\epsilon_k + \alpha+ \beta{\Phi_1}^2) \delta\Psi_0\\
    (\epsilon_k + \alpha+ \beta{\Phi_1}^2-2\beta_s{\Phi_1}^2 )\delta\Psi_{-1}
    \end{pmatrix}
\end{equation}

It is easy to get the frequency of $\delta\Psi_1$:
\begin{equation}
\hbar \omega_1 = \sqrt{(\epsilon_k + \alpha+ \beta{\Phi_1}^2)^2
    + 2\beta{\Phi_1}^2 (\epsilon_k + \alpha+ \beta{\Phi_1}^2) } .
\end{equation}
Under BEC, $f^0_b$ satisfies the relation $\partial f^0_b/\partial
\Phi_1=0$ , which yields
\begin{equation}\label{eq33}
\alpha+ \beta{\Phi_1}^2=0.
\end{equation}
This condition guarantees a gapless Goldstone mode in the
Bogliubov form. We can also derive frequencies of $\delta\Psi_0$
and $\delta\Psi_{-1}$:
\begin{equation}\label{eq34}
\hbar \omega_0 = \epsilon_k + \alpha+ \beta{\Phi_1}^2,~~~~
\hbar\omega_{-1} = \epsilon_k + \alpha+ \beta{\Phi_1}^2-
2\beta_s{\Phi_1}^2  .
\end{equation}
Under condition of Eq. (\ref{eq33}), $\omega_0$ also becomes
gapless. Up to now, the results of {Refs. \protect\cite{ho}} and
{\protect\cite{ohmi}} are reproduced, except that the meaning of
parameters is different. One advantage of the phenomenological
approach is that it could reveal more evidently the relation
between the Goldstone mode and the phase transition.

However, we should note that the above derivation is not
self-consistent, because it neglects the magnetization in the
normal part of the gas. According to discussions in Section 3, the
FM transition takes place above BEC. So near the BEC temperature,
the magnetization in normal gas is much larger than in the
condensate. Apparently, this FM phase in the thermal cloud
(described by $f_m$) should be taken into account. In this case,
the coupling $f_c$ between the two phases plays an important role.
$f_c$ consists of two terms,
\begin{subequations}
\begin{eqnarray}
f^0_c &=& -g M_0{\Phi_1}^2 ,\\
f^I_c &=& -g M_0(\delta\Psi^*_1\delta\Psi_1
               -\delta\Psi^*_{-1}\delta\Psi_{-1}) \nonumber\\
    &&-\frac {\sqrt{2}}2 g \Phi_1 (\delta\Psi^*_0{\delta M_+}
               + \delta\Psi_0{\delta M_-}).
\end{eqnarray}
\end{subequations}
The self-consistent solution to phase transitions and Goldstone
modes in spinor condensates should be acquired by treating the
total free energy $f_t=f_m+f_b+f_c$ as a whole\cite{Gu3}. The
obtained results suggested that the Bogliubov mode remains
unchanged, but spin waves in the thermal cloud and in the
condensate are coupled together,
\begin{eqnarray}
f^I_{sf} = \begin{pmatrix}{\delta M_+} & \delta\Psi^*_0 \end{pmatrix}
    \begin{pmatrix}ck^2+\frac {g{\Phi_1}^2}{2M_0} &
      -\frac {\sqrt{2}}2 g \Phi_1 \\  -\frac {\sqrt{2}}2 g \Phi_1
      & \epsilon_k +gM_0  \end{pmatrix}
    \begin{pmatrix}{\delta M_-} \\ \delta\Psi_0\end{pmatrix}
\end{eqnarray}
This equation indicates that the transverse spin fluctuations in
both phases are {\it gapped} solo. But after considering the
coupling, the Goldstone theorem is recovered. The spectrum for
this coupled mode at long wave length is given by
\begin{equation}
\hbar \omega_0 \approx \frac {gM_0ck^2+\frac
{g{\Phi_1}^2}{2M_0}\epsilon_k}
           {gM_0-\frac {g{\Phi_1}^2}{2M_0}}
      \approx ck^2+\frac 12 \frac {\Phi_1^2}{M_0^2}\epsilon_k.
\end{equation}
Once again, we obtain a $k^2$-formed spectrum for spin waves.

Many attempts have been made to evaluate excitation spectra at
finite temperatures microscopically, on the basis of perturbation
theory\cite{szepfalusy,ohtsuka} and generalized self-consistent
Hartree-Fock theory\cite{szirmai}. The obtained results show that
the dispersion relations are in the same forms as at zero
temperature. The phonon velocity of the Bogliubov mode decreases
while the spin-wave stiffness $c_s$ increases as the temperature
is increasing. But these theories are not very much applicable
near the BEC temperature, not only because of the invalidity of
the calculating method itself at the temperature regime under
discussion, but also because of the neglect of the normal FM
phase.

As a comparison to the ferromagnetic condensate, let us take a
glance at the spin waves in antiferromagnetic Bose gases in which
the Bose-Einstein condensate is in the ``polar" state: only
spin-$0$ bosons condense. In this case the spin and density waves
are described by $\delta
n=\sqrt{n_0}(\delta\Psi_0+\delta\Psi^*_0)$, $\delta M_-
=\sqrt{n_0}(\delta\Psi_{-1}+\delta\Psi^*_1)$ and $\delta M_+
=\delta M_-^\dag$\cite{ho,ohmi}. The spectra are
\begin{equation}
\hbar\omega_0 = \sqrt{\epsilon_k^2 +
      2\frac {4\pi \hbar^2 }{m}g_0n_0\epsilon_k},~~~~
\hbar\omega_{\pm} = \sqrt{\epsilon_k^2 +
      2\frac {4\pi \hbar^2 }{m}g_sn_0\epsilon_k} .
\end{equation}
At very small $k$, the spin-wave spectrum $\hbar \omega_{\pm} \sim
\sqrt{g_s}k$. The dispersion relation is linear in $k$ which
coincides with that in a Heisenberg antiferromagnet in which
$\hbar\omega \sim |J|k$. But, the dependence of the spin wave
velocities on the couplings are different.

\section{Conclusions and Further discussions}
In summary, some theoretical results concerning ferromagnetic
transitions in spinor Bose gases are reviewed. The subject covers
the origin of the coupling (which induces the transition), the
transition itself (including the critical point and its relation
with BEC), and the Goldstone mode (as a result of the transition).
A unique feature of the FM transition in bosonic systems is that
it takes place always above Bose-Einstein condensation, regardless
of the magnitude of the coupling. The spectrum of the spin wave
takes the same form as in a conventional ferromagnet, $\omega\sim
k^2$, while the spin-wave stiffness coefficient consists of two
different parts: one coming from the thermal cloud, the other from
the Bose condensate, which embodies the ``two-fluid" feature of
the system.

The above theoretical results are devoted mainly to a homogeneous
system in the thermodynamic limit and the actual experimental
situations for ultracold atomic gases are not well taken into
account. In the following we discuss briefly some experimental
facts which may cause significant differences.

1) {\it The trapping potential}. Atomic gases are experimentally
confined by a trapping potential which can be approximated as
being harmonic. One direct consequence of the confinement is that
it changes the density of state of the gas and thus the critical
behaviors\cite{silvera}. So how the trapping potential affects on
the phase diagram of the FM spinor boson deserves further study.
More recently, Huang {\it et al.}\cite{wjhuang} studied BEC in
trapped $F=1$ spinor bosons with FM couplings, but the FM
transition was not considered.

2) {\it The number of confined atoms}. At present, the number of
confined atoms is typically in the range of $10^6-10^8$, far away
from the thermodynamic limit. So questions arise: do theories
derived in the thermodynamic limit hold in this case? Does the
concept of spontaneous symmetry breaking apply? And to what extent
does it apply? The observation of BEC suggests that the
spontaneous symmetry breaking is still a valid concept, and the
observed critical points agree with theories quite
well\cite{gerbier}. But it is still an open question whether it is
so for the FM transition.

3) {\it The spin conservation rule}. It is observed that the total
spin approximately conserves through the evaluation of the
Bose-Einstein condensate\cite{ketterle,schma,chang}. From Ref.
\protect\cite{schma,chang}, this conservation dominates the spin
dynamics and the final states of the system. This is in conflict
with the picture of spontaneous symmetry breaking which says the
total spin does not conserve in the ground state. It might be a
consequence of the departure from the thermodynamic limit.

Isoshima {\it et al.} studied phase transitions in $F=1$ spinor
bosons under the spin conservation rule\cite{isoshima1}. They
argued that the system may have two spatially phase-separated Bose
condensates with $m_F=1$ and $-1$ respectively.

Although many work has been done, the research on ferromagnetism
in spinor bosons is still in a very primary stage. Many questions
remain open. Direct comparison between theories and experiments
is far from being reached.

\section*{Acknowledgements}
The author gratefully acknowledges collaboration with K.
Sengstock, K. Bongs, R. A. Klemm, helpful discussions with K.
Scharnberg and D.I. Uzunov, and support from the Deutsche
Forschungsgemeinschaft through the Graduiertenkolleg
``Spectroscopy of localized atomic systems", No. 463.

\end{document}